# Unveiling the Origin of the Insulating Ferromagnetism in LaMnO$_3$ Thin Film

Y. S. Hou, H. J. Xiang[*], and X. G. Gong[*]

Key Laboratory of Computational Physical Sciences (Ministry of Education), State Key Laboratory of Surface Physics, and Department of Physics, Fudan University, Shanghai 200433, People's Republic of China

By combining genetic algorithm optimizations, first-principles calculations and the double-exchange model studies, we have unveiled that the exotic insulating ferromagnetism in LaMnO$_3$ thin film originates from the previously unreported G-type $d_{3z^2-r^2}/d_{x^2-y^2}$ orbital ordering. An insulating gap opens as a result of both the orbital ordering and the strong electron-phonon coupling. Therefore, there exist two strain induced phase transitions in the LaMnO$_3$ thin film, from the insulating A-type antiferromagnetic phase to the insulating ferromagnetic phase and then to the metallic ferromagnetic phase. These phase transitions may be exploited in tunneling magnetoresistance and tunneling electroresistance related devices.

Perovskite LaMnO$_3$ (LMO), a fundamental strongly correlated electron system and the parent compound of the colossal magnetoresistance manganites, displays a complex correlation among structural, orbital, magnetic, and electronic degrees of freedom and has been extensively studied over the past decades [1]. Recently, LMO thin films attract considerable attentions not only because many interesting and emerging phenomena are discovered in LMO thin film related superlattices [2-5], but also because LMO thin films display exotic behavior different from bulk. Experiments discovered a surprising insulating ferromagnetic phase in LMO thin films epitaxially grown on the square-lattice SrTiO$_3$ (STO) substrate [2, 7-13]. The origin of this puzzling insulating FM phase is unclear. The unexpected ferromagnetism was considered to be extrinsic and deficiency of La [10, 14] was suggested as a possible explanation to it. However, it is not consistent with the fact that FM phase in the LMO thin films with cation deficiency tends to be metallic [7, 15, 16]. Previous models [17, 18] and recent first-principles studies [19, 20] on LMO thin films predicted the metallic FM phase, instead of the experimentally observed insulating FM phase. It is important to address how the prototypical antiferromagnetic Mott insulator transforms to an insulating ferromagnet.

To answer such an intriguing question, we carry out a comprehensive theoretical study on the LMO thin film by combining genetic algorithm (GA) optimizations, first-principles calculations and the orbital-degenerate double-exchange (DE) model studies. Our extensive GA simulations give an insulating FM phase of the LMO thin film, which crystallizes in the monoclinic *P2₁/n* structure. This monoclinic *P2₁/n* structure has a peculiar arrangement pattern of Jahn-Teller (JT) distortions of MnO$_6$ octahedra, giving rise to the previously unreported three-dimensionally $d_{3z^2-r^2}/d_{x^2-y^2}$ -alternated orbital order. According to the Goodenough-Kanamori rules [21], this kind of orbital orders induces three-dimensional ferromagnetism. First-principles calculations show that the band gap of the monoclinic *P2₁/n* FM phase is 0.16 eV, compatible with the experimentally observed excitation energy [2] ($\approx 0.14 \, \text{eV}$). By virtue of the orbital-degenerate DE model, it is revealed that the band gap opens as a result of both the orbital ordering and the strong electron-phonon coupling. We also show that LMO thin film transforms from the insulating A-type antiferromagnetic (A-AFM) phase to the insulating ferromagnetic phase, and then to the metallic ferromagnetic phase when the lateral lattice constant decreases.

In this work, the widely adopted global optimization technique GA [22-25] is used to search the ground state structure of the LMO thin film. The GA method we employed here is similar to that described in Ref. [25] with the exception that the spin information of the magnetic ions is explicitly kept in the crossover operation and a mutation operation related to the spin direction is also added. A $\sqrt{2} \times \sqrt{2} \times 2$ perovskite supercell containing four formula units (20 atoms) is used. In GA searches, in order to match the experimentally epitaxial strain induced by the square-lattice substrate STO, the two in-plane lattice vectors are fixed at $\boldsymbol{a}_{LMO} = \boldsymbol{b}_{LMO} = \sqrt{2}\boldsymbol{a}_{STO}$ where $\boldsymbol{a}_{STO}$ is the lattice constant (3.905 Å) of the cubic STO substrate. However, both the length and the direction of the out-of-plane lattice vector $\boldsymbol{c}_{LMO}$ and the internal ionic coordinates are fully optimized. First-principles calculations based on density functional theory are performed using the generalized gradient approximation with the PW91 parameterization plus on-site Coulomb repulsion U method (GGA+U) as implemented in *Vienna Ab Initio Simulation Package* (VASP) [26-30]. Because the experimentally measured [31] $U = 3.5$ eV describes well the bulk LMO (see Sec. I of the Supplemental Material) [32] and gives a band gap (0.16 eV) of the LMO thin film close to the experimentally observed excitation energy [2] ($\approx 0.14$ eV), it is applied to Mn 3$d$ electrons in the present work. A further detailed study showed that our main results remain correct if a reasonable U value is used (see Sec. II of the Supplemental Material) [32].

Our calculations show that, *Pbnm* LMO thin film strained on STO is metallic FM, consistent with the recent work of Lee *et al*. [20]. By extensive GA searches, however, we find an insulating FM phase, which has a total energy lower than the metallic *Pbnm* FM phase by 6.9 meV/f.u. (Fig. 1). This insulating FM phase crystallizes in a slightly distorted monoclinic *P21/n* structure with $\beta = 90.74^0$. Two kinds of Mn ions, denoted as Mn-A and Mn-B, are found in the monoclinic *P21/n* structure (Fig. 2b) while all Mn ions are equivalent in the orthorhombic *Pbnm* structure (Fig. 2a). Mn-A has a MnO$_6$ octahedron elongated along the *c* axis (Fig. 2c) while Mn-B has a MnO$_6$ octahedron stretched in the *ab* plane (Fig. 2d). What is most significant is that these two different kinds of Mn atoms are arranged in a checkerboard G-type manner (Fig. 2b). This is rather different from the bulk LMO where the relevant $Q_3$ modes [1] of the JT distortions of all MnO$_6$ octahedra are with their principal axes lying within the *ab* plane and these axes are alternatively arranged in this plane.

The band structure of the majority spin (up) of the monoclinic *P21/n* FM phase is plotted in Fig. 3a. Obviously, the monoclinic *P21/n* FM phase is insulating. It has an indirect band gap of 0.16 eV, compatible with the experimentally observed excitation energy [2] ($\approx 0.14\,\text{eV}$). Moreover, its bandwidth is narrower than that of the metallic *Pbnm* FM phase (Fig. 3c). We expect that the insulating *P21/n* FM phase can be confirmed by the angle resolved photoemission spectroscopy (ARPES) experiment.

We now examine the orbital order in the insulating *P21/n* FM phase. Since the MnO$_6$ octahedron of Mn-A is elongated along the *c* axis, the $d_{3z^2-r^2}$ orbital is lower in energy than the $d_{x^2-y^2}$ orbital and thus the single $e_g$ electron occupies the $d_{3z^2-r^2}$ orbital, which can be evidenced by the PDOS (Fig. 4a). The notable split between the peak of the $d_{3z^2-r^2}$ orbital and that of the $d_{x^2-y^2}$ orbital is consistent with the large JT distortion [1] $Q_A = \sqrt{Q_2^2 + Q_3^2} = 0.27$ Å. In contrast, the orbital occupation is just opposite for Mn-B. The $d_{x^2-y^2}$ orbital is lower in energy as a result of the in-plane stretch of MnO$_6$ octahedron and thus the single $e_g$ electron mainly occupies the

$d_{x^2-y^2}$ orbital, which can also be verified by the PDOS (Fig. 4b). Likewise, its relative weak split between the peak of $d_{3z^2-r^2}$ orbital and that of $d_{x^2-y^2}$ orbital is due to the small JT distortion $Q_B = \sqrt{Q_2^2 + Q_3^2} = 0.14$ Å. Fig. 3b shows the $e_g$ charge density integrated from -1.5 eV to the Fermi level, which displays the orbital order of the insulating *P21/n* FM phase. In this energy interval, the spectral density of Mn-A has predominantly $d_{3z^2-r^2}$ character while Mn-B has predominantly $d_{x^2-y^2}$ character. Therefore, we firstly report that the insulating *P21/n* FM phase has a type of three-dimensionally $d_{3z^2-r^2}/d_{x^2-y^2}$-alternated orbital order, which is different from the $d_{3z^2-r^2} + d_{x^2-y^2}$ type [18] (Fig. 3d) and completely different from the $d_{3x^2-r^2}/d_{3y^2-r^2}$ type [33].

Based on the established orbital order, the magnetic interactions among the Mn atoms in the insulating *P21/n* FM phase can be deduced according to the Goodenough-Kanamori rules [21]. As a result of the three-dimensionally $d_{3z^2-r^2}/d_{x^2-y^2}$-alternated orbital order, the half-filled $\sigma$-bond $d_{3z^2-r^2}$ orbital from Mn-A overlaps with the empty $\sigma$-bond $d_{3z^2-r^2}$ orbital from Mn-B along the cubic [001] axis through the middle $O-2p_z$ orbital, giving rise to strong ferromagnetic interactions. Besides, the half-filled $\pi$-bond $t_{2g}$ orbitals will give rise to an antiferromagnetic interaction between Mn-A and Mn-B. Since the overlap of the $\sigma$-bonding electrons is greater than that of the $\pi$-bonding electrons, the ferromagnetic interaction turns out to be stronger than the antiferromagnetic one. Therefore the superexchange interaction between Mn-A and Mn-B along the cubic [001] axis is ferromagnetic. This mechanism also applies to the magnetic interactions between Mn-A and Mn-B in the *ab* plane, where the mutually interacting orbitals are the half-filled $\sigma$-bonding $d_{x^2-y^2}$ orbital of Mn-B and the empty $\sigma$-bonding $d_{x^2-y^2}$ orbital of Mn-A. To sum up, Mn-A and Mn-B interact ferromagnetically along

[100], [010] and [001] axes (see Sec. III of the Supplemental Material) [32]. The deduced magnetic interactions among Mn atoms are verified by density functional theory (DFT) calculations by means of the four-states mapping method [34]. The considered nearest neighbor (NN) magnetic interaction paths $J_{cc}$, $J_{ab1}$ and $J_{ab2}$ are shown in Fig. 2b. As expected, our calculations find $J_{cc} = 9.73$ meV, $J_{ab1} = 3.91$ meV and $J_{ab2} = 5.94$ meV. All of them are ferromagnetic. The next nearest neighbor (NNN) magnetic interactions are found to be much weaker than the NN ones. Therefore, the magnetic ground state is ferromagnetic. With the DFT calculated magnetic exchange constants, our MC simulations (see Sec. IV of the Supplemental Material) [32] lead to a transition temperature $T_C = 446$ K, higher than experimentally measured ones [7, 11] ranging from 115K to 240K. A possible explanation to this discrepancy is that the epitaxial strain is gradually relaxed from the substrate to the film surface, thus the FM phase locates just near the strained film-substrate interface while the bulk-like A-AFM phase dominates at the film surface, similar to the case of the strain-induced ferromagnetism in the antiferromagnetic LuMnO$_3$ thin film [35].

In order to unveil the mechanism of the insulating ferromagnetism, we employ the orbital-degenerate DE model with one $e_g$ electron per Mn$^{3+}$ ion. An infinite Hund coupling limit [1], i.e., $J_H = \infty$, is adopted in our study. With this useful simplification, the DE model Hamiltonian reads

$$H = -\sum_{\langle ij \rangle}^{\alpha\beta} t_{\alpha\beta}^{\bar{a}} \left( \Omega_{ij} d_{i\alpha}^+ d_{j\beta} + H.C. \right) + \sum_{\langle ij \rangle} J_{AF}^{\bar{a}} \vec{S}_i \cdot \vec{S}_j + \lambda \sum_i \left( -Q_{1i} n_{1i} + Q_{2i} \tau_{xi} + Q_{3i} \tau_{zi} \right)$$
$$+ \frac{1}{2} \sum_i \left( 2Q_{1i}^2 + Q_{2i}^2 + Q_{3i}^2 \right) \qquad (1).$$

A detailed description of this Hamiltonian is given in the Sec. V of Supplementary Materials [32]. Note that the second term, the NN antiferromagnetic superexchange interaction between Mn $t_{2g}$ spins, is left out of considerations in the present work since both the *P2$_1$/n* and the *Pbnm* phases are FM. The DFT relaxed structures of the

*P21/n* FM and the *Pbnm* FM phases are used. Because the breathing modes ($Q_1$) are much smaller than the JT modes ($Q_2$ and $Q_3$) in both phases, they are ignored in our study. Lastly $t_0 = 0.52\,\text{eV}$ and $\lambda = 1.4$ ($\lambda$ usually estimated [1] between 1.0 and 1.6) are used [36, 37]. This set of parameters results in an energy difference between the *P21/n* FM and the *Pbnm* FM phases close to that obtained from the DFT calculations. By exactly diagonalizing the DE Hamiltonian, we show the band structures of the *P21/n* FM and *Pbnm* FM phases in Fig. 3a and Fig. 3c, respectively. As expected, the band structure of *P21/n* FM phase is characteristic of insulator and that of *Pbnm* FM phase is characteristic of metal. The profiles of both band structures obtained from the model well reproduce that obtained from the DFT calculations, indicating that the parameters involved in the model are appropriately selected.

Now let's address why the *P21/n* FM phase is insulating while the *Pbnm* FM phase is metallic. For the *P21/n* FM phase, as the three-dimensionally $d_{3z^2-r^2}/d_{x^2-y^2}$-alternated orbital order has almost no overlap along the *c* axis ($t_{ab}^z = t_{ba}^z = 0$) between the occupied orbitals, its bandwidth $W_{P21/n}$ is determined mainly by the in-plane overlap ($t_{ab}^x = t_{ba}^x = -\frac{\sqrt{3}}{4}t_0$ or $t_{ab}^y = t_{ba}^y = \frac{\sqrt{3}}{4}t_0$) between the lower-energy occupied orbital $d_{3z^2-r^2}$ of Mn-A and the lower-energy occupied orbital $d_{x^2-y^2}$ of Mn-B. So its bandwidth $W_{P21/n}$ is proportional to $\sqrt{3}t_0$. Likewise, the bandwidth $W_{Pbnm}$ of *Pbnm* FM phase determined by the $d_{3z^2-r^2} + d_{x^2-y^2}$-type orbital order is proportional to $3t_0$. Thus, *P21/n* FM phase has a narrower bandwidth than *Pbnm* FM phase. On the other hand, the onsite energy splitting of the two Mn$-e_g$ orbitals due to the electron-phonon coupling in an isolated MnO$_6$ octahedron is proportional to the strength of the JT distortions [19]:

$$\epsilon_\pm = \pm\lambda\sqrt{Q_2^2 + Q_3^2} \qquad (2).$$

Thus, the onsite energy splitting in *P21/n* FM phase is much larger than that in *Pbnm*

FM phase since the JT distortions in the former are much severer than that in the latter. Therefore, *P21/n* FM phase possesses a finite band gap as a result of the narrow bandwidth and the large on-site $e_g$ level splitting (Fig. 4a and Fig. 4b). For the *Pbnm* FM phase, however, its bandwidths of Mn $e_g$ bands are so wide and the electron-phonon coupling is so weak that no band gap opens (Fig. 4c). Furthermore, decompositions of the total energies of both phases (Table I) show that, the electron-phonon coupling causes a much larger energy lowering in the *P21/n* FM phase than the *Pbnm* FM phase although *P21/n* FM phase has higher hopping energy and lattice elastic energy than the *Pbnm* FM phase. In conclusion, it is revealed that the orbital order and the electron-phonon cooperatively make the *P21/n* FM phase insulating while the *Pbnm* FM phase metallic and that the electron-phonon coupling plays a vital role in stabilizing the insulating *P21/n* FM phase as the ground state of the LMO thin film strained on STO.

Finally, using the same strategy applied to the LMO thin film strained on STO, we have systematically investigated the property dependence of LMO thin films on lattice constants, which can be experimentally tuned by selecting different square-lattice substrates. Here, only the *P21/n* and *Pbnm* space groups are considered. As usual, several common magnetic orders in perovskite are considered, namely, FM, A-AFM, C-type AFM (C-AFM) and G-type AFM (G-AFM) spin orderings. The results are shown in Fig. 1. As the C-AFM and G-AFM spin orders do not appear as the lowest energy phase in the considered lattice constant window, they are not shown in Fig. 1 and not further discussed. Our results show that LMO thin films with lattice constant ranging from 3.88 Å to 4.03 Å are insulating FM with the monoclinic *P21/n* structure and the three-dimensionally $d_{3z^2-r^2}/d_{x^2-y^2}$ -alternated orbital order. However, LMO thin films with lattice constants smaller than 3.88 Å are metallic FM with the *Pbnm* structure and $d_{3z^2-r^2}+d_{x^2-y^2}$ -type orbital order. When the lattice constant is larger than 4.04 Å, the ground state of the LMO thin films becomes the *Pbnm* structure with an insulating A-AFM order and $d_{3x^2-r^2}/d_{3y^2-r^2}$ -type orbital order. The metallic *Pbnm* FM and the insulating *Pbnm* A-AFM phases are consistent with the work of Lee *et al.* [20] on LMO thin film with large compressive and tensile epitaxial strain. The phase transitions from the insulating *P21/n* FM (*P21/n* FM I) phase to the metallic *Pbnm* FM (*Pbnm* FM M) phase and then to the insulating *Pbnm* A-AFM (*Pbnm* A-AFM I) phase are intuitively illustrated by the obvious discontinuity of the lattice constant $c$, shown in the upper panel of Fig. 1. The phase diagram shown in Fig. 1 is also confirmed by our GA optimizations.

In summary, the physical origin of the well-known and puzzling insulating ferromagnetism experimentally observed in the LaMnO$_3$ thin film grown on the square-lattice SrTiO$_3$ substrate has been investigated. We find that the insulating ferromagnetic phase is intrinsically from strain induced orbital ordering, instead of extrinsic reasons such as defects. It crystallizes in a monoclinic *P21*/n structure, which has two different kinds of MnO$_6$ octahedra: one is elongated along the $c$ axis and the other one is stretched in the $ab$ plane. They are arranged in a checkerboard G-type manner, giving rising to a previously unreported three-dimensionally $d_{3z^2-r^2}/d_{x^2-y^2}$ − alternated orbital order, which naturally leads to the ferromagnetism. Double Exchange model reveals that the band gap opens due to both the orbital ordering and the strong electron-phonon coupling. Finally, we find that epitaxially strained LaMnO$_3$ thin film transforms from the insulating A-type antiferromagnetic phase to the insulating ferromagnetic phase, and then to the metallic ferromagnetic phase when the lateral lattice constant decreases. If LMO thin film is epitaxially grown on some specified piezoelectric materials, an electric-field-induced metal-insulator transition and an electric field control of the magnetism can be realized experimentally at $Pbnm\,\text{FM M} \to P21/n\,\text{FM I}$ and $P21/n\,\text{FM I} \to Pbnm\,\text{A}-\text{AFM I}$ phase boundaries, respectively. These electric-field induced phase transitions may be exploited in tunneling magnetoresistance (TMR) and tunneling electroresistance (TER) related devices.

Work was partially supported by NSFC, the Special Funds for Major State Basic Research, Foundation for the Author of National Excellent Doctoral Dissertation of

China, The Program for Professor of Special Appointment at Shanghai Institutions of Higher Learning, Research Program of Shanghai municipality and MOE. We thank for H. R. Liu, X. Gu and J. H. Yang, X. F. Zhai and S. Dong for valuable discussions.

**Figures**

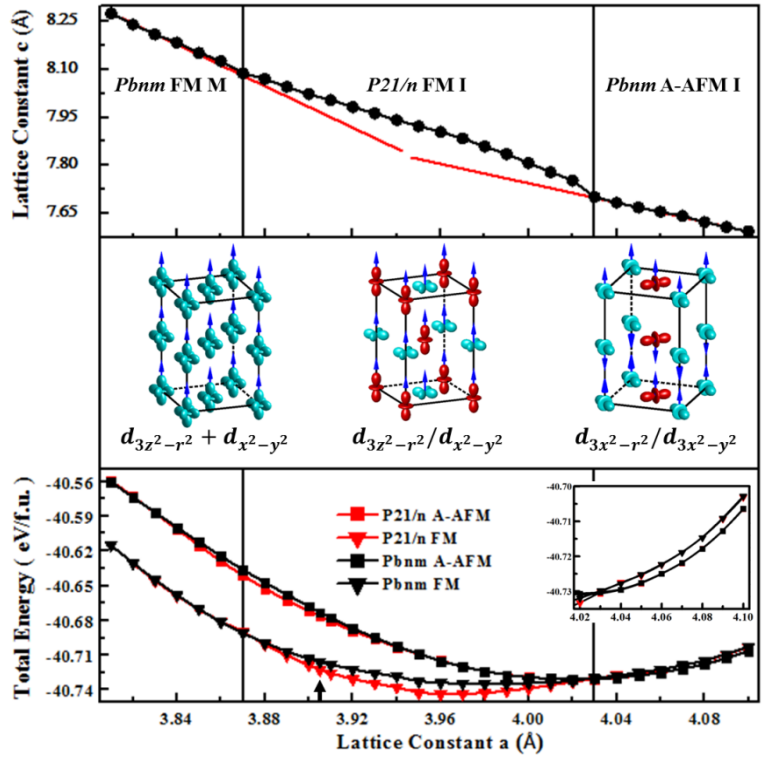

Figure 1 (Color online) Calculated Phase diagram of LaMnO$_3$ thin films. Within the considered lattice constant window, LMO thin films transform from metallic (M) *Pbnm* FM to insulating (I) *P21/n* FM, then to the insulating *Pbnm* A-AFM phases. LMO thin films with the nearest neighbor Mn-Mn distance close (or equal ) to that of bulk LMO [38] ($a = 3.99$ Å) do not take the bulk structure because in this case LMO thin film is slightly stretched along the *a* axis and seriously compressed along the *b* axis. The evolution of the lattice constant *c* of the ground state is given in the upper panel. The two red straight lines are guides to the eyes. The magnetic and orbital orders of each ground state are depicted in the middle panel. Blue arrows represent spins. Bottom panel gives the total energies of the *P21/n* A-AFM, *P21/n* FM, *Pbnm* A-AFM and *Pbnm* FM phases as a function of the square-lattice substrate lattice constant *a*. Black arrow indicates the lattice constant of the SrTiO3 substrate. Note that *P21/n* FM phase merges with the *Pbnm* FM phase when the lattice constant becomes smaller than 3.88 Å and that *P21/n* structure merges with the *Pbnm* structure when the lattice constant becomes larger than 4.03 Å.

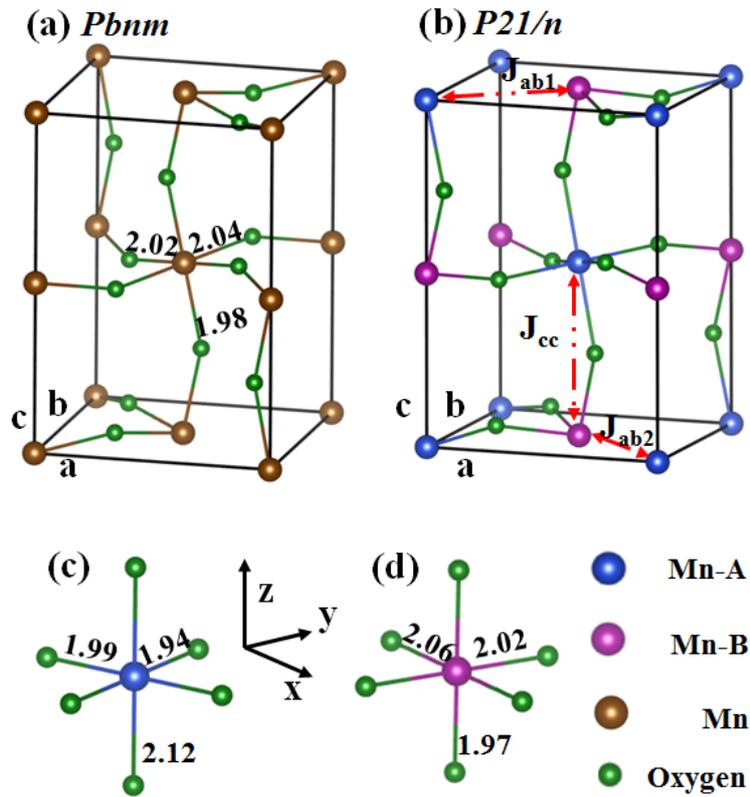

Figure 2 (Color online) Structure of LaMnO$_3$ thin films strained on SrTiO$_3$. (a) Structure of the metallic *Pbnm* FM phase. (b) Structure and magnetic interaction paths $J_{cc}$, $J_{ab1}$ and $J_{ab2}$ of the insulating *P2$_1$/n* FM phases. MnO$_6$ octahedra of Mn-A (c) and Mn-B (d) of the insulating *P2$_1$/n* FM phase are shown. Their local coordinate systems shown between (c) and (d) are chosen in such a way that the local *z* axis is nearly along the *c* axis and *xy* plane nearly in the *ab* plane. Crystallographic axes are given by a, b and c. All numbers give Mn-O bond lengths in units of Å.

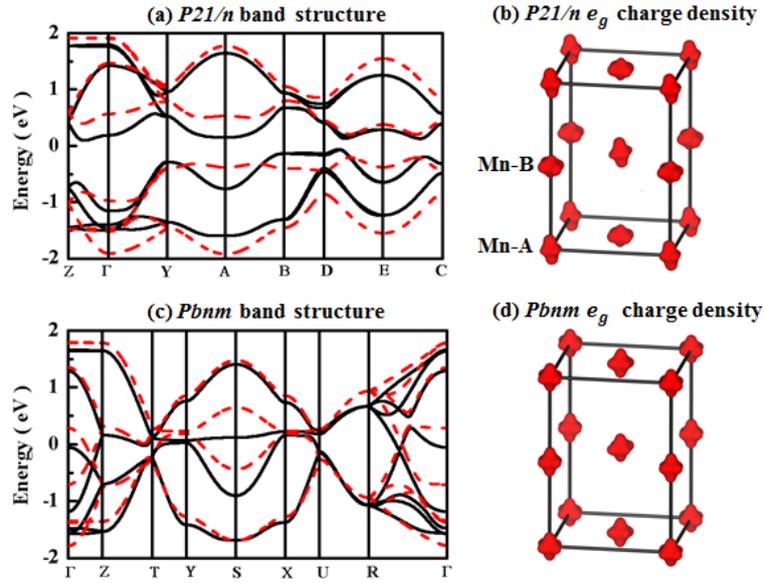

Figure 3 (Color online) Band structure and charge density of LaMnO$_3$ film strained on SrTiO$_3$. Band structures of the insulating *P21/n* FM and the metallic *Pbnm* FM phases are shown in (a) and (c), respectively. Black solid (red dashed) lines in both (a) and (c) are from the DFT calculations (orbital-degenerate double-exchange model). The Fermi level is set to zero. DFT calculated $e_g$ charge density distributions of the insulating *P21/n* FM and the metallic *Pbnm* FM phases in the energy window of 1.5 eV widths just below the Fermi level are plotted in (b) and (d), respectively.

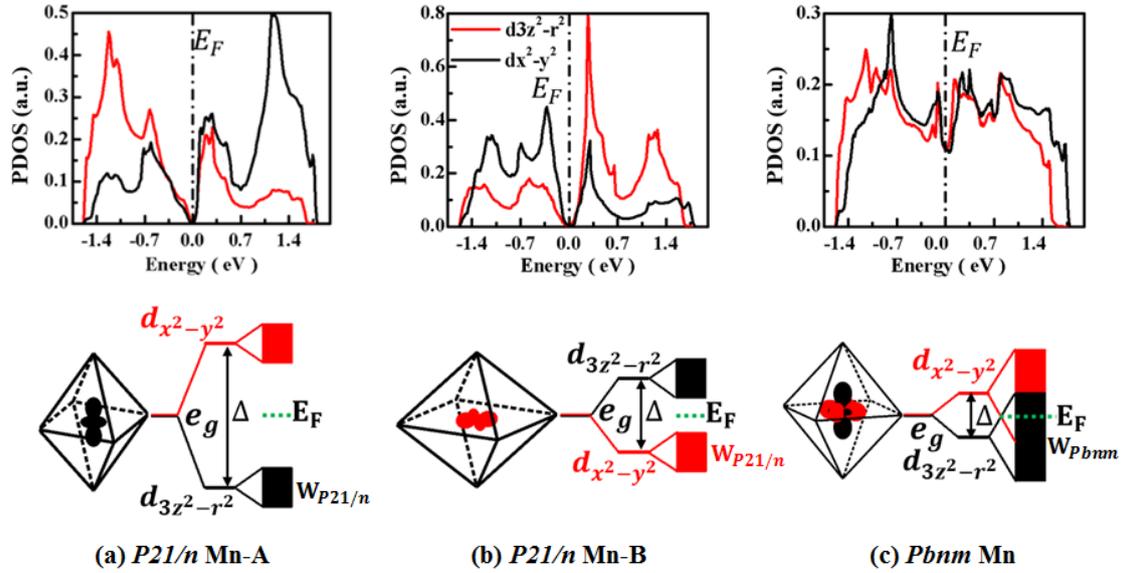

Figure 4 (Color online) Partial density of states (PDOS) and onsite energy splitting of LaMnO$_3$ thin film strained on SrTiO$_3$. PDOS of Mn-A, Mn-B of the insulating *P21/n* FM phase and the Mn atom of the metallic *Pbnm* FM phase are shown in the upper panels of the (a), (b) and (c), respectively. The Fermi level is indicated by the black vertical lines and set to be zero. Diagrammatic sketches of the onsite energy splitting of the two $Mn-e_g$ orbitals shown in bottom panels of (a), (b) and (c) correspond to Mn-A, Mn-B of the insulating *P21/n* FM phase and the Mn atom of the metallic *Pbnm* FM phase, respectively. The parameter $\Delta$ is the energy difference between the $d_{3z^2-r^2}$ and the $d_{x^2-y^2}$ orbitals. The black (red) rectangular box represents the band derived from the $d_{3z^2-r^2}$ ($d_{x^2-y^2}$) orbital and the green horizontal lines indicate the Fermi level.

Table I. Decompositions of the total energies ($E_{tot}$) of the insulating *P21/n* FM and the metallic *Pbnm* FM phases (20 atoms, namely, four formula units) into the hopping energy ($E_{hopping}$), electron-phonon coupling ($E_{ele-ph}$) and lattice elastic energy ($E_{lattice}$).

|  | $E_{hopping}$ (eV) | $E_{ele-ph}$ (eV) | $E_{lattice}$ (eV) | $E_{tot}$ (eV) |
|---|---|---|---|---|
| *P21/n* | -2.4354 | -1.5878 | 0.9045 | -3.1187 |
| *Pbnm* | -3.0319 | -0.1978 | 0.1332 | -3.0966 |

Supplementary Materials for

Unveiling the Origin of the Insulating Ferromagnetism in LaMnO$_3$ Thin Film

Y. S. Hou, H. J. Xiang*, and X. G. Gong*

Key Laboratory of Computational Physical Sciences (Ministry of Education), State Key Laboratory of Surface Physics, and Department of Physics, Fudan University, Shanghai 200433, P. R. China

1. A detailed study on the bulk LaMnO$_3$

   With the experimentally measured [1] $U = 3.5$ eV and $J = 0.9$ eV applied to the Mn 3$d$ electrons, our full structural optimizations successfully reproduce the A-type AFM (A-AFM) magnetic ground state in the bulk LaMnO$_3$ (LMO) with a lower total energy than the FM state by amount of 8.5 meV/formula unit (f.u.), although it is not a trivial task to achieve this within DFT framework [2]. The calculated direct gap is 1.2 eV, consistent with the optical measurements [3]. Besides, the relevant magnetic exchange constants are calculated to be $J_{ab} = 2.09$ meV and $J_c = -1.13$ meV, respectively. They are consistent with experimentally measured [4] $J_{ab} = 1.85$ meV and $J_c = -1.1$ meV. With these calculated magnetic exchange constants, Monte Carlo (MC) simulations (Fig. S1) reveal that the transition temperature $T_N = 116$ K, well consistent with the experimentally observed one [5] $T_N = 140$ K. Intriguingly, our systematical studies (Fig. S2) clearly show that either too small or too large $U$ cannot correctly obtain the A-AFM magnetic ground state. We find that $U$ should be between 2.5 eV and 4.0 eV to reproduce the A-AFM ground state.

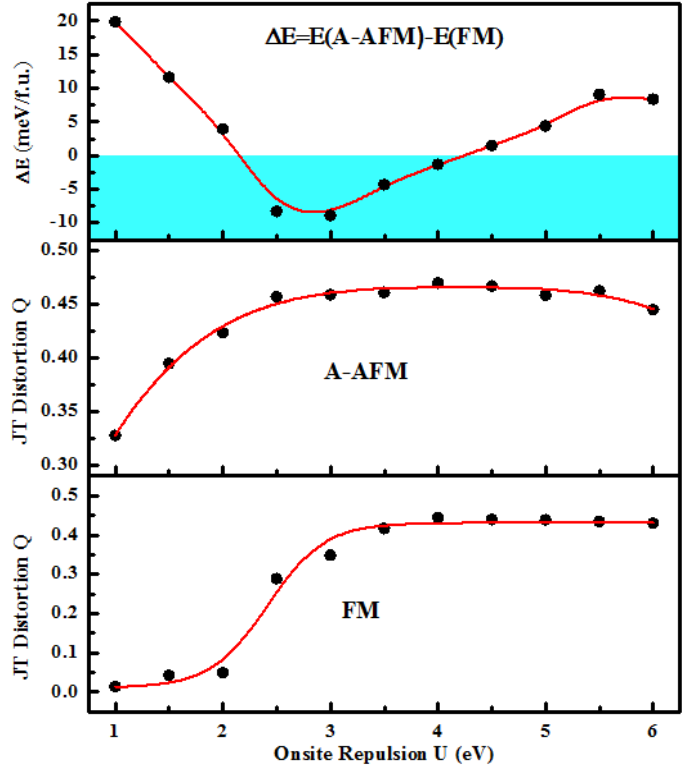

Fig. S1 The dependence of the energy difference between the A-AFM and FM (upper panel), the Jahn-Teller distortions of the A-AFM state (middle panel) and the FM state (bottom panel) on the U. Jahn-Teller Q is defined as $Q = \sqrt{Q_2^2 + Q_3^2}$.

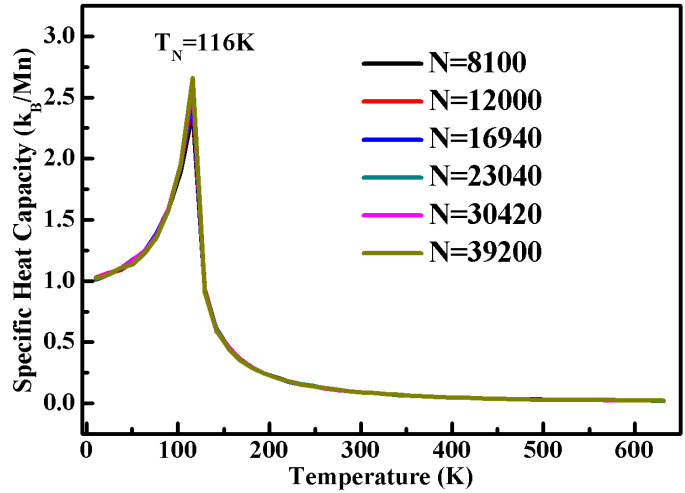

Fig. S2 Specific heat capacity of bulk LaMnO$_3$ as a function of temperature from MC simulations with different lattice sizes (up to 39200 sites). A sharp $\lambda$ peak of the specific heat capacity locates at $T_N = 116$ K.

2. A detailed study on the impact of $U$ upon the epitaxially strained LaMnO$_3$ film with SrTiO$_3$ lattice constant

Detailed study show that the monoclinic *P21/n* state is always more stable than the *Pbnm* state when a reasonable $U$ between 2.5 and 4.0 eV that describes well bulk LaMnO$_3$ is adopted (Fig. S3). It is found that for $U$ less than 2.0 eV, the monoclinic *P21/n* state cannot be stabilized because the small $U$ significantly underestimates the JT distortions. In other words, only the *Pbnm* structure is obtained. For the $U$ larger than 3.0 eV, the *P21/n* FM phase becomes insulating and has an appreciably lower total energy than the metallic *Pbnm* FM phase. Moreover, both the band gap of the insulating *P21/n* FM phase and the total energy difference between the insulating *P21/n* FM and the metallic *Pbnm* FM phases increase with $U$. It is worthy of noting that any $U$ ranging from 1.0 eV to 6.0 eV cannot open a band gap in the *Pbnm* FM phase.

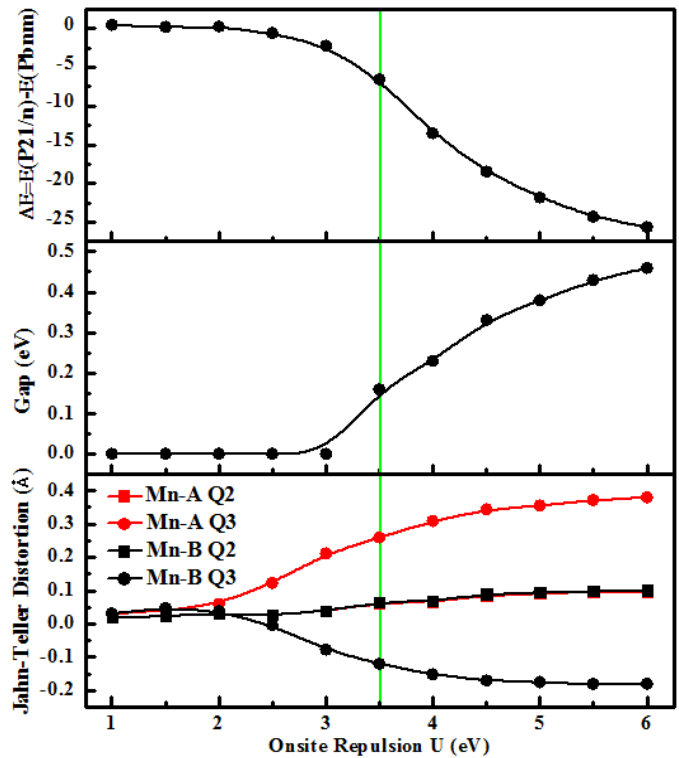

Fig. S3 Dependence of the relative stability between the *P21/n* FM and the metallic *Pbnm* FM phases, the band gap and the Jahn-Teller distortions of the *P21/n* FM phase on the $U$.

3. Origin of ferromagnetism in the insulating *P21/n* phase

In order to investigate the magnetic interactions in the insulating *P21/n* phase, ferromagnetic (FM) and antiferromagnetic (AFM) spin configurations between Mn-A and Mn-B (Fig. S4) are considered. If Mn-A and Mn-B ferromagnetically interact along the [001] axis, the occupied $d_{3z^2-r^2}$ orbital of Mn-A can strongly hybridize ($\sigma$-type) with the unoccupied $d_{3z^2-r^2}$ orbital of Mn-B, and thereby the total energy is lowered significantly (upper panel of Fig. S4). However, if they interact antiferromagnetically, the hybridization between the occupied $d_{3z^2-r^2}$ orbital of Mn-A and the unoccupied $d_{3z^2-r^2}$ orbital of Mn-B are almost negligible because these two orbital has a large energy difference due to the Hubbard repulsion $U$ (upper panel of Fig. S4). Besides, the hybridizations ($\pi$-type) between $t_{2g}$ orbitals of Mn-A and Mn-B are also negligible. Thus, there is only a tiny energy gain for the antiferromagnetically interacting case. Therefore, we find that the FM interaction between Mn-A and Mn-B along the [001] axis is preferred to the AFM interaction. A similar mechanism applies to the magnetic in-plane interactions between Mn-A and Mn-B except that the interacting orbitals are the unoccupied $d_{x^2-y^2}$ orbital of Mn-A and the occupied $d_{x^2-y^2}$ orbital of Mn-B (bottom panel of Fig. S4). To sum up, Mn-A and Mn-B interact ferromagnetically along [100], [010] and [001] axes.

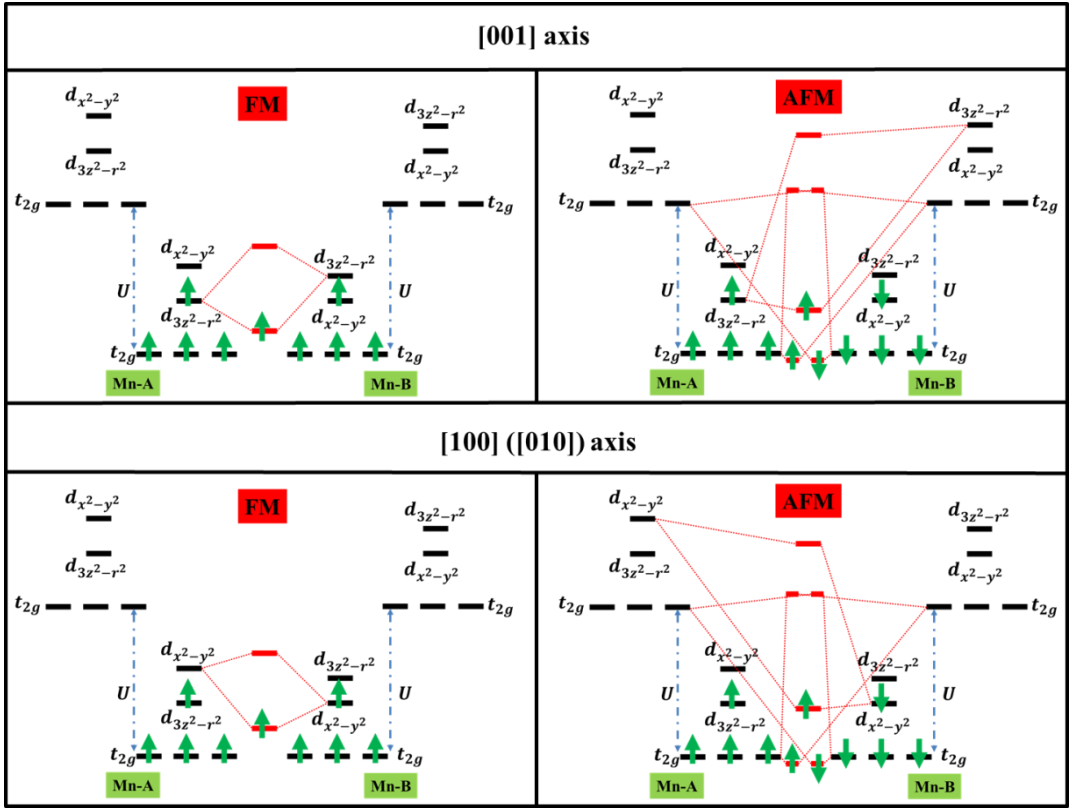

Fig. S4 The orbital interactions between Mn-A and Mn-B with FM and AFM spin configurations, respectively. Upper panel is for the Mn pair along the [001] axis while the bottom panel is for the Mn pair along the [100] and [010] axes.

4. Monte Carlo simulation of the transition temperature $T_C$ of the epitaxially strained LaMnO$_3$ film with SrTiO$_3$ lattice constant

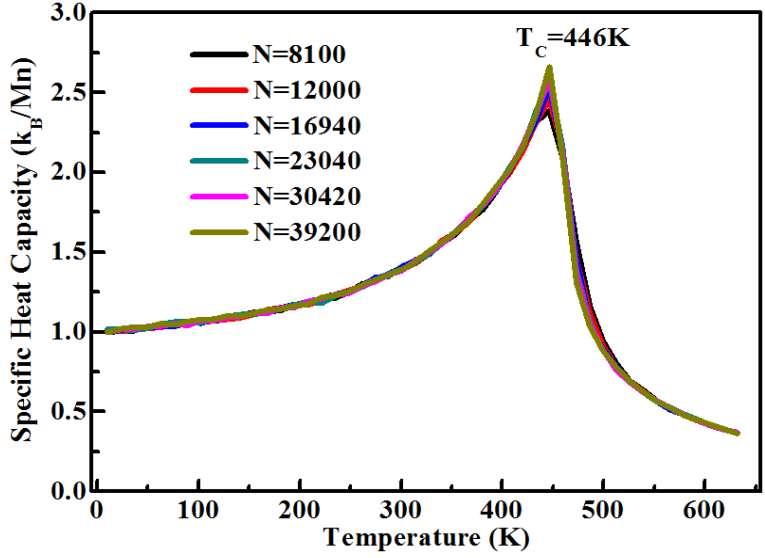

Fig. S5 Specific heat capacity of LaMnO$_3$ thin film grown on STO as a function of temperature from MC simulation with different lattice sizes (up to 39200 sites). A sharp $\lambda$ peak of the specific heat capacity locates at $T_C = 446$ K.

5. A detailed description of the orbital-degenerate double-exchange model Hamiltonian

The model Hamiltonian in our present work reads

$$H = -\sum_{<ij>}^{\alpha\beta} t_{\alpha\beta}^{\vec{a}} \left( \Omega_{ij} c_{i\alpha}^{+} c_{j\beta} + H.c. \right) + \sum_{<ij>} J_{AF}^{\vec{a}} \vec{S_i} \cdot \vec{S_j} + \lambda \sum_i \left( -Q_{1i} n_i + Q_{2i} \tau_{xi} + Q_{3i} \tau_{zi} \right)$$
$$+ \frac{1}{2} \sum_i \left( 2Q_{1i}^2 + Q_{2i}^2 + Q_{3i}^2 \right) \tag{1},$$

where $d_{i\alpha}^{+}$ ($d_{i\alpha}$) is the creation (annihilation) operator for the $e_g$ electron on the orbital $\alpha = |x^2 - y^2>(a)$ and $\beta = |3z^2 - r^2>(b)$, with its spin parallel to the localized $t_{2g}$ spin $\vec{S_i}$; $\vec{a} = \mathbf{x}, \mathbf{y}, \mathbf{z}$ is the direction of the link connecting the two nearest neighbor (NN) Mn$^{3+}$ sites; Berry phase $\Omega_{ij} = \cos\frac{\theta_i}{2}\cos\frac{\theta_j}{2} + \sin\frac{\theta_i}{2}\sin\frac{\theta_j}{2} e^{-i(\phi_i - \phi_j)}$ arises due to the infinite Hund coupling, where $\theta$ and $\phi$ are the polar and azimuthal angles of the $t_{2g}$ spins, respectively. In the model Hamiltonian, the first term is the standard DE interaction. The hopping parameters are $t_{aa}^x = -\sqrt{3}t_{ab}^x = -\sqrt{3}t_{ba}^x = 3t_{bb}^x = \frac{3t_0}{4}$, $t_{aa}^y = \sqrt{3}t_{ab}^y = \sqrt{3}t_{ba}^y = 3t_{bb}^y = \frac{3t_0}{4}$, $t_{aa}^z = t_{ab}^z = t_{ba}^z = 0$ and $t_{bb}^z = t_0$. The second term is the NN $t_{2g}$ spins interaction through the antiferromagnetic superexchange $J_{AF} > 0$. The third term is the electron-phonon coupling, where $\lambda$ is a dimensionless constant and the $e_g$-orbital operators are $n_i = d_{ia}^+ d_{ia} + d_{ib}^+ d_{ib}$, $\tau_{xi} = d_{ia}^+ d_{ib} + d_{ib}^+ d_{ia}$, and $\tau_{zi} = d_{ia}^+ d_{ia} - d_{ib}^+ d_{ib}$. The last term is the lattice elastic energy. JT modes ($Q_2$ and $Q_3$) and breathing mode ($Q_1$) are defined in Ref. [6].